\documentstyle[preprint,eqsecnum,aps]{revtex}

\tightenlines

\begin{document}

\draft

%\preprint{}

\title{Improved lattice QCD with quarks: the 2 dimensional case}

\author{Jun-Qin Jiang$^{b}$,  Xiang-Qian Luo$^{a,c,d}$,  
Zhong-Hao Mei$^{c,d}$,
Hamza Jirari$^{e}$,  Helmut Kr\"oger$^{e}$,   Chi-Min Wu$^{f}$}

\address{$^a$CCAST(World Laboratory), P.O. Box 8730, Beijing 100080, 
People's Republic of China\\
$^b$Department of Physics, Guangdong Institute of Education, 
Guangzhou 510303, China\\
$^c$Department of Physics, Zhongshan University, Guangzhou 510275, 
China\footnote{Official and mailing Address. 
E-mail: stslxq@zsulink.zsu.edu.cn}\\
$^d$ Center for Computational Physics, School of Physics Science and Engineering, 
Zhongshan University, Guangzhou 510275, China\\
$^e$ D\'epartement de Physique, Universit\'e Laval,
Qu\'ebec, Qu\'ebec G1K 7P4, Canada\\
$^{f}$ Institute for High Energy Physics, Academia Sinica, 
Beijing 100039, China}

\date{\today}

\maketitle

\begin{abstract}
QCD in two dimensions is investigated using the
improved fermionic lattice
Hamiltonian proposed by Luo, Chen, Xu, and Jiang.
We show that the improved theory leads to a significant reduction of 
the finite lattice spacing errors. 
The quark condensate and the mass of lightest 
quark and anti-quark bound state in the strong coupling phase 
(different from t'Hooft phase) are computed.
We find agreement between our results and the analytical ones
in the continuum.
\end{abstract}


\narrowtext 

\section{INTRODUCTION}
\label{sec1}

Wilson's lattice formulation of QCD is the most reliable and powerful
non-perturbative approach to the physics of strongly interacting particles, 
but its progress was hampered by systematical errors mainly 
due to the finite value of the lattice spacing $a$.

Wilson's gluonic action differs from the continuum Yang-Mills action 
by order of $O(a^2)$, while the error of Wilson's quark action is bigger, 
i.e., the error being of the order $O(a)$.
Due to quantum effects, there is an additional problem called ``tadpole".
If $a$ and the bare coupling constant $g$ were small enough, these errors
would be negligible. Unfortunately, even in recent 4-dimensional
lattice QCD simulations on the most powerful computers, 
the lattice spacing $a$ is larger than $0.1fm$ and 
the lattice coupling $g_{lat}$ is bigger than $0.9$.  For these lattice parameters,
violation of scaling is still obvious and
the extrapolation of the results to the $a \to 0$ limit induces 
unknown systematic uncertainties when extracting continuum physics.

One of the most efficient ways for reducing these systematic errors is 
the Symanzik improvement program \cite{Symanzik83}: adding local, nearest neighbor 
or next-nearest-neighbor interaction terms to the Wilson lattice theory
so that the finite $a$ errors become higher order in $a$. 
During recent  years, 
application of the Symanzik program has been a major topic.
There have been several proposals on this subject:

\noindent
(a) For the gauge sector, Lepage proposed a tadpole improved action 
\cite{Lepage96}, reducing the errors from $O(a^2)$ to $O(a^4)$.
Luo, Guo, Kr\"oger, and Sch\"utte 
constructed a tadpole improved Hamiltonian \cite{Luo98_1,Luo98_2} with
the same accuracy.

\noindent
(b) For the fermionic sector, Hamber and Wu
(one author of the present paper)
proposed the first improved action \cite{Hamber83}, 
by adding next-nearest-neighbor 
interaction terms to the Wilson quark action 
so that the $O(a)$ error is canceled. There have been some
numerical simulations \cite{Lombardo93,Borici97,Fiebig97}
of hadron spectroscopy using the Hamber-Wu action.
Sheikholeslami and Wohlert \cite{Sheik85}, L\"uscher and Weisz \cite{Luscher85}
worked on fermionic improvement and recent progress has been achieved on non-perturbative improvement via implementation of PCAC by the ALPHA collaboration \cite{Luscher98}.
Luo, Chen, Xu, and Jiang (two authors of the present paper)
constructed an improved Hamiltonian \cite{Luo94}, 
which was tested in the Schwinger model ($\rm{QED_2}$). 
There are other possibilities 
for improving Wilson's quark theory \cite{Luo96_1}.
The purpose of this work is to demonstrate numerically that fermionic improvement works also in the Hamiltonian formulation for the case of QCD in two dimensions.

Like Quantum Electrodynamics in 2 dimensions ($\rm{QED_{2}}$)
also Quantum Chromodynamics in 2 dimensions ($\rm{QCD_2}$) 
has the properties of chiral symmetry breaking, an anomaly and confinement.
However, the latter has gluonic self interactions, which makes it more similar 
to QCD in 4 dimensions ($\rm{QCD_4}$) 
than to the Schwinger model.  
In addition to having confinement and chiral-symmetry breaking, 
$\rm{QCD_2}$ has much richer spectrum 
because of the non-abelian gauge interactions 
between ``gluons" and ``quarks". 
Early in 1974, 't Hooft \cite{tHooft74} did 
pioneering work on $\rm{QCD_2}$ using the $1/N_C$ expansion (where $\rm{QCD}$ corresponds to the gauge symmetry group $\rm{SU(N_C)}$).
He showed in the limit $N_{C} \to \infty$ that planar diagrams are 
dominant and they can be summed up in a bootstrap equation. From that he obtained a meson-like mass spectrum. 
This model has been extensively studied in the large $N_{C}$ limit and also in the chiral limit $m_{q} \to 0$ (where $m_{q}$ is the free quark mass). Zhitnitsky \cite{kn:Zhit85} has pointed out that there two distinct phases: \\
(i) $N_{C} \to \infty$ firstly, and then $m_{q} \to 0$ afterwards, \\
(ii) $m_{q} \to 0$, and then $N_{C} \to \infty$. \\
The first case corrersponds to $g << m_{q}$, which is the weak coupling regime.
It descibes the phase considered by t`Hooft, 
where the following relation holds,
\begin{equation}
N_{C} \to \infty, ~~~ g^{2}N_{C} = \mbox{const.} ~~~ 
m_{q} >> g \sim {1 \over \sqrt{N_{C}}}.
\end{equation}
t'Hooft found a spectrum of states (numbered by $n$) given by
\begin{equation} 
M_{n} \sim \pi^{2} m_{0} n, ~~~ m_{0}^{2} = {g^{2}N_{C} \over \pi}.
\end{equation}
Zhitnitsky \cite{kn:Zhit85} has computed the gluon condensate and the quark condensate. The quark condensate is given by
\begin{equation}
\langle \bar{\psi} \psi \rangle = - N_{C} \sqrt{{g^{2}N_{C} \over 12 \pi}}.
\label{eq:Zhit}
\end{equation}
The second case corresponds to $g >> m_{q}$ which is the strong coupling regime.
In the strong coupling regime the observed spectrum is different \cite{kn:Balu80,kn:Bhat82,kn:Stei80}.
Steinhardt \cite{kn:Stei80} has computed the masses of the following particles: soliton (baryon), anti-soliton (anti-baryon) and soliton-antisoliton (baryon-antibaryon) bound state. Bhattacharya \cite{kn:Bhat82} has shown that 
in the chiral limit there are free fields of mass
\begin{equation} 
M = g \sqrt{{N_{C}+1 \over 2\pi}}.
\label{eq:Bhat}
\end{equation}
Note that for $N_C=1$ this coincides with the Boson mass in the Schwinger model.
Grandou et al. \cite{kn:Gran88} have obtained quark condensate in the chiral limit (for arbitrary $g$) and obtained 
\begin{equation}
\langle \bar{\psi} \psi \rangle \sim - g N_{C}^{3/2}
\end{equation}
which confirms Zhitnitsky`s result obtained in the weak coupling regime.

\bigskip

Of course, $\rm{QCD_2}$ is much simpler than $\rm{QCD_4}$. 
It has been used to mimic the properties of $\rm{QCD_4}$ 
such as vacuum structure, hadron scattering and decays, 
and charmonium picture.
Unlike the massless Schwinger model, unfortunately, $\rm{QCD_2}$ 
is no longer exactly solvable. 

The first lattice study of $\rm{QCD_{2}}$ was done by Hamer \cite{kn:Hame82} 
using Kogut-Susskind fermions. 
He computed for $\rm{SU(2)}$ the mass spectrum in the t'Hooft phase.
In 1991, Luo $et~al.$ \cite{Luo91_1} 
performed another lattice field theory study of this model 
($\rm{SU(2)}$ and $\rm{SU(3)}$) using Wilson fermions.
As is shown later, the results for lattice $\rm{QCD_2}$ 
with Wilson quarks were found
to be strongly dependent on the unphysical Wilson parameter $r$, 
the coefficient of the $O(a)$ error term.
The purpose of this paper is to show that in the case of 
$\rm{QCD}_2$, the improved theory \cite{Luo94} 
can significantly reduce these errors.

The remaining part of the paper is organized as follows.
In Sect. \ref{sec2}, we review some features of the Hamiltonian
approach as well as the improved Hamiltonian
for quarks proposed in Ref. \cite{Luo94}. 
In Sect. \ref{sec3}, the wave functions of 
the vacuum and the vector meson are constructed, 
and the relation between the continuum chiral condensate and
lattice quark condensate is developed.
The results for the quark condensate and the mass spectrum 
are presented in Sect. \ref{sec4} and discussions 
are presented in Sect. \ref{sec5}.

\section{IMPROVED HAMILTONIAN FOR QUARKS}
\label{sec2}

Although numerical simulation in the Lagrangian formulation has become
the main stream and a lot of progress has been made over the last 
two decades, there are areas where progress has been quite slow and new
techniques should be explored:
for example, computation of the $S$-matrix and cross sections,
wave functions of vacuum, hadrons and glueballs,
QCD at finite baryon density, or the computation of 
QCD structure functions in the region of small $x_{B}$ and $Q^{2}$. 
In our opinion the Hamiltonian approach is a viable alternative 
\cite{Kroger92,Guo97} and
some very interesting results 
\cite{Luo96_2,Luo97,Luo98_3,Briere89,Kroger97,Kroger98,Schutte97}
have recently been obtained.
Many workers in the Lagrangian formulation nowadays have followed 
ideas similar to the Hamiltonian approach 
(where the time is continuos, i.e., $a_t=0$)
by considering anisotropic lattices 
with lattice spacings $a_t << a_s$.
The purpose is to improve the
signal to noise ratio in the spectrum computation \cite{Lepage96}. 

In the last ten years, we have done  a lot of work
\cite{Luo91_1,Luo89,Luo90_1,Luo90_2,Chen90,Luo91_2,Luo92}
on Hamiltonian lattice field theory, where the conventional 
Hamiltonian in the Wilson fermion sector is used:
\begin{eqnarray*}
H_{f}= H_{m} + H_{k} + H_{r},
\end{eqnarray*}
\begin{eqnarray*}
H_{m} = m\sum_{x} \bar {\psi}(x) \psi (x),
\end{eqnarray*}
\begin{eqnarray*}
H_{k}={1 \over 2a} \sum_{x,k} \bar{\psi}(x) \gamma_{k} U(x,k) \psi (x+k),
\end{eqnarray*}
\begin{eqnarray}
H_{r} = {r \over 2a} \sum_{x,k}[\bar{\psi}(x) \psi (x)-
\bar{\psi} (x) U(x,k) \psi (x+k)],
\label{unimprovedH}
\end{eqnarray}
where $a$ is now the spacial lattice spacing $a_s$, 
$U(x,k)$ is the gauge link variable at site 
$x$ in the direction $k= \pm j$ ($j$ is the unit vector),
and $\gamma_{-j} = -\gamma_{j}$, $H_m$, $H_k$, $H_r$ are 
respectively the mass term, kinetic term and Wilson term. 
The Wilson term ($r \not= 0$), proportional to $O(ra)$, 
is introduced to avoid the fermion species doubling, with the price of
explicit chiral-symmetry breaking even in the vanishing bare fermion mass limit.
As discussed in Sec. \ref{sec1}, 
the $O(a)$ error in $H_{f}$ indeed
leads to lattice artifacts if $a$ or $g_{lat}$ is not small enough.

Similar to the Hamber-Wu action \cite{Hamber83}, 
where some next-nearest-neighbor interaction terms 
are added to the Wilson action to cancel the $O(ra)$ error, 
we proposed a $O(a^2)$ improved Hamiltonian in Ref. \cite{Luo94}: 
\begin{eqnarray*}
H_{f}^{improved}= H_{m} + H_{k}^{improved} + H_{r}^{improved},
\end{eqnarray*}
\begin{eqnarray*}
H_{k}^{improved}= 
{b_{1} \over 2a} \sum_{x,k}\bar{\psi} (x) \gamma_{k} U(x,k) \psi (x+k)
\end{eqnarray*}
\begin{eqnarray*}
+{b_{2} \over 2a}\sum_{x,k}\bar{\psi} (x) \gamma_{k} U(x,2k) \psi (x+2k),
\end{eqnarray*}
\begin{eqnarray*}
H_{r}^{improved}={r \over 2a} \sum_{x,k}\bar{\psi}(x) \psi (x)
\end{eqnarray*}
\begin{eqnarray*}
-c_{1}{r \over 2a} \sum_{x,k}\bar{\psi} (x) U(x,k) \psi (x+k)
\end{eqnarray*}
\begin{eqnarray}
-c_{2}{r \over 2a} \sum_{x,k}\bar{\psi} (x) U(x,2k) \psi (x+2k).
\label{ImprovedH}
\end{eqnarray}
Here $U(x,2k)=U(x,k)U(x+k,k)$ and the coefficients 
$b_{1}, b_{2}, c_{1}$  and $c_{2}$ are given by
\begin{eqnarray}
b_{1}={4 \over 3},  b_{2}=-{1 \over 6}, 
c_{1}={4 \over 3}, c_{2}=-{1 \over 3}.
\end{eqnarray}
These coefficients are the same for any d+1 dimensions
and gauge group. 
The results shown in this paper correspond to this set of parameters. 
However, the following set of paramters
\begin{eqnarray}
b_{1}={1},  b_{2}=0, 
c_{1}={4 \over 3}, c_{2}=-{1 \over 3},
\end{eqnarray}
where only the Wilson term is improved, gives very similar results.
With the absence of the $O(ra)$ error, we
expect that we can extract the continuum physics in a more reliable way.

Lattice $\rm{QCD_2}$ in the Hamiltonian approach 
has some nice features which
simplify the computations considerably:

\noindent
(a) The magnetic interactions are absent and the tadpole factor $U_0=1$
due to the fact that $U_p=U^{\dagger}_p=1$. Therefore, there is only
color-electric energy term in the gluonic Hamiltonian:
\begin{eqnarray}
H_{g} = {g^{2} \over 2a} \sum_{x,j} E_{j}^{\alpha}(x) E_{j}^{\alpha}(x),
\end{eqnarray}
with $j=\vec{1}$ and $\alpha=1, ..., N_c^{2}-1$.

\noindent
(b) The quantum (tadpole) effects, if any, are highly suppressed 
in the fermionic sector as $O(g_{lat}^2a) \to O(a^3)$ 
and in the gluonic sector as $O(g_{lat}^2a^2) \to O(a^4)$. 
The reason is the super-renormalizability of 1+1 dimensional
theories, where 
\begin{eqnarray}
g_{lat}=ga. 
\end{eqnarray}

All this means that
in 1+1 dimensions, classical improvement of the fermionic
Hamiltonian is sufficient.
In Ref. \cite{Luo94}, the mass spectrum of the Schwinger model 
was used to test the improved program.
In the following sections, we will provide evidence in $\rm{QCD_2}$
to support the efficiency of the improved Hamiltonian.
The results for the quark condensate 
and mass spectrum further confirm
our expectation.

\section{VACUUM AND VECTOR PARTICLE STATES}
\label{sec3}

The vacuum wave function
is constructed in the same way as in \cite{Luo94}:
\begin{eqnarray}
\vert \Omega \rangle =exp(iS)\vert 0 \rangle,
\end{eqnarray}
where
\begin{eqnarray*}
S=\sum_{n=1}^{N^S_{trun}} \theta_{n}S_n
\end{eqnarray*}
\begin{eqnarray*}
S_1 = i \sum_{x,k}\psi^{\dagger} (x) \gamma_{k} 
U(x,k) \psi (x+k)
\end{eqnarray*}
\begin{eqnarray*}
S_2 = i \sum_{x,k}\psi^{\dagger} (x) \gamma_{k} U(x,2k) \psi (x+2k) 
\end{eqnarray*}
\begin{eqnarray*}
S_3= i \sum_{x,k}\psi^{\dagger} (x) \gamma_{k} U(x,3k) \psi (x+3k)
\end{eqnarray*}
\begin{eqnarray}
... 
\label{Vacuum}
\end{eqnarray}
with $\theta_n$
determined by minimizing the vacuum 
energy. $\vert 0 \rangle$ is the bare vacuum defined by
\begin{eqnarray}
\xi(x) \vert 0\rangle = \eta(x) \vert 0\rangle = 
E^{\alpha}_{j}(x) \vert 0\rangle =0.
\end{eqnarray}
Here $\xi$ and $\eta^{\dagger}$ are respectively the up and down 
components of the $\psi$ field.
Such a form of the fermionic vacuum (\ref{Vacuum}) 
has also been discussed extensively
in the literature 
\cite{Luo91_1,Luo89,Luo90_1,Luo90_2,Chen90,Luo91_2,Luo92}.

One is interested in computing the wave function of the lowest lying energy state
which is the flavor-singlet vector meson $\vert V \rangle$, 
similar to the case of the Schwinger model. 
The wave function is created by
a superposition of some operators $V_{n}$ with the given quantum numbers 
\cite{Luo94,Fang92}
\begin{eqnarray*}
V_{0}=i \sum_{x}\bar{\psi} (x) \gamma_{1} \psi (x),
\end{eqnarray*}
\begin{eqnarray*}
V_{1}=i \sum_{x,k}\bar{\psi} (x) \gamma_{1} U(x,k) \psi (x+k),
\end{eqnarray*}
\begin{eqnarray*}
V_{2}=i \sum_{x,k}\bar{\psi} (x) \gamma_{1} U(x,2k) \psi (x+2k),
\end{eqnarray*}
\begin{eqnarray*}
V_{3}=i \sum_{x,k}\bar{\psi} (x) \gamma_{1} U(x,3k) \psi (x+3k),
\end{eqnarray*}
\begin{eqnarray}
... 
\label{Operator}
\end{eqnarray}
acting on the vacuum state $\vert \Omega \rangle$, i.e.,
\begin{eqnarray}
\vert V \rangle = \sum_{n=0}^{N^V_{trun}} A_{n} 
[V_{n}-\langle \Omega \vert V_{n} \vert \Omega\rangle] 
\vert\Omega \rangle.
\label{Vector}
\end{eqnarray}
The criterion for choosing the truncation orders
$N^S_{trun}$ and $N^V_{trun}$ is the convergence of the
results.
An estimate for the vector mass $M_V$
is the lowest eigenvalue $~ ~ min(E_V) ~ ~$ of the following equations
\begin{eqnarray*}
\sum_{n_1=0}^{N^V_{trun}}(H^{V}_{n_{2}n_{1}}-E_{V}U^{V}_{n_{2}n_{1}})A_{n_1}=0,
\end{eqnarray*}
\begin{eqnarray}
det \vert H^{V}-E_{V}U^{V} \vert =0,
\label{Eigen}
\end{eqnarray}
where the coefficients $A_{n_1}$ 
are determined by solving the equations, and
the matrix elements $H^{V}_{n_{2}n_{1}}$ and $U^{V}_{n_{2}n_{1}}$
are defined by 
\begin{eqnarray*}
H^{V}_{n_{2}n_{1}}
=\langle V_{n_{2}}\vert [ H^{improved}_{f}+H_g ] \vert  V_{n_{1}}
\rangle ^{L},
\end{eqnarray*}
\begin{eqnarray}
U^{V}_{n_{2}n_{1}}=\langle V_{n_{2}}\vert  V_{n_{1}}\rangle^{L}.
\label{Matrix}
\end{eqnarray}
Here the superscript $L$ means 
only the matrix elements proportional to the
lattice size $L$ are retained (those proportional to higher order of $L$ do not contribute).
Detailed discussions can be found in \cite{Fang92}. We can estimate the continuum
$M_V$ from
\begin{eqnarray}
{M_V\over g}={aM_V\over g_{lat}},
\label{Mscaling}
\end{eqnarray}
if the right hand side is independent of $g_{lat}$.

In lattice field theory with Wilson fermions,
chiral symmetry is explicitly broken even in the
bare vanishing mass limit. Therefore, the fermion condensate
$\langle \bar{\psi} \psi \rangle_{free}$
is non-vanishing and should be subtracted in a way described in
Refs. \cite{Luo91_1,Chen90,Luo91_2,Luo90_3}:
\begin{eqnarray}
\langle \bar{\psi} \psi \rangle_{sub}=
\langle \bar{\psi} \psi \rangle - \langle \bar{\psi} \psi \rangle_{free},
\end{eqnarray}
where for one flavor
\begin{eqnarray*}
\langle \bar{\psi} \psi \rangle
= {1 \over LN_c} \langle \Omega \vert \bar{\psi} \psi
\vert \Omega \rangle_{m=0}
\end{eqnarray*}
\begin{eqnarray*}
\langle \bar{\psi} \psi \rangle_{free}
= {1 \over LN_c} {\partial E_{\Omega_{free}} \over \partial m} \vert_{m=0}
\end{eqnarray*}
\begin{eqnarray*}
=- \int_{-\pi/a}^{\pi/a} d pa
\lbrace [ {r \over a} (1-c_{1}cos pa- c_{2}cos 2pa)]^{2}
\end{eqnarray*}
\begin{eqnarray*}
+[{sin pa \over a} (b_{1}+b_{2}cos pa)]^{2} \rbrace
\end{eqnarray*}
\begin{eqnarray*}
%\over
/\lbrace [ {r \over a} (1-c_{1}cos pa- c_{2}cos 2pa) ]^{2}
\end{eqnarray*}
\begin{eqnarray}
+[{sin pa \over a} (b_{1}+b_{2}cos pa)]^{2} \rbrace^{1/2}.
\end{eqnarray}

From  $\langle \bar{\psi} \psi \rangle_{sub}$, we can get an estimate of the
continuum quark condensate $\langle \bar{\psi} \psi \rangle_{cont}$
\begin{eqnarray}
{\langle \bar{\psi} \psi \rangle_{cont} \over g}= 
{\langle \bar{\psi} \psi \rangle_{sub} \over g_{lat}},
\label{scaling}
\end{eqnarray}
if the right hand side does not depend on the bare coupling $g_{lat}$.

It is well known that spontaneous chiral-symmetry breaking
originates from the axial anomaly and
there is no Goldstone pion in 
quantum field theory in 1+1 dimensions.
Therefore, for  Wilson fermions, one can not fine-tune the
$(r,m)$ parameter space as in $\rm{QCD}_4$ to reach the chiral limit.
However, the chiral limit can be approximated as the $m \to 0$ limit
as long as the lattice spacing error is sufficiently small. 
For Wilson fermions, as discussed in Ref. \cite{Azcoiti96},
this is not well justified for finite $g_{lat}$ and $a$ 
because of the $O(ra)$ error,
but for the improved theory, as shown in Refs. \cite{Borici97,Luo94},
this would be a reasonably good approximation
since the chiral-symmetry breaking term
is much smaller, i.e., $O(ra^2)$.

\section{RESULTS FOR QUARK CONDENSATE AND VECTOR MASS}
\label{sec4}

To increase the accuracy of the techniques described in Sect. \ref{sec3},
we include higher order contributions in Eq. (\ref{Vacuum})
than those in Ref. \cite{Luo94} so that better convergence of the results
can be obtained. 

Figure 1 and Figure 3 show 
$\langle \bar{\psi} \psi \rangle_{sub}/g_{lat}$ as a function of $1/g_{lat}^{2}$ 
in 2-dimensional lattice SU(2) and SU(3), respectively, gauge theories with Wilson fermions.
Figure 5 and Figure 7 show 
$aM_V/g_{lat}$ as a function of $1/g_{lat}^{2}$ 
in 2-dimensional lattice SU(2) and SU(3), respectively, gauge theories with Wilson fermions.
As one can see, the results for $r=1$ deviate obviously from
those for $r=0.1$, 
which is attributed to the $O(ra)$ error of the Wilson term.

The corresponding results from the Hamiltonian with improvement 
are presented 
in Figure 2, Figure 4, Figure 6, and Figure 8.
One observes that the differences between the
results for $r=1$ and $r=0$ are significantly reduced.
Most impressively, the data for the quark condensate
coincide each other.
A similar $r$ test has also been used in \cite{Alford97}
for checking the efficiency of the improvement program.

To get an idea about how $\rm{QCD}_2$ behaves for large $N_C$,
we have computed the quark condensate $\langle \bar{\psi} \psi \rangle$ in the chiral limit for $N_C=3,4,5,6$. This is shown in Figure 9.
We have compared our numerical results with the theoretical result by Zhitnitsky \cite{kn:Zhit85}, Eq.(\ref{eq:Zhit}). Although Zhitnitsky's result was obtained in the weak coupling phase, it qualitatively agrees with the strong coupling result of Ref.\cite{kn:Gran88}. Remarkably, our results agree very well with 
Zhitnitsky's weak coupling result.  
Secondly, we show in Figure 10 the lowest lying mass of the mass spectrum again for $N_C=3,4,5,6$ in the chiral limit. 
Note that this particle corresponds to the vector particle in the Schwinger model. In $\rm{QCD_2}$ it corresponds to free particles (the mass of which remains finite when $m_{q}$ goes to zero) distinct from 
``sine-Gordon" soliton particles (the mass of which goes to zero when $m_{q}$ goes to zero, see \cite{kn:Stei80}). We compare our numerical result with the analytical strong coupling result by Bhattacharya \cite{kn:Bhat82}, Eq.(\ref{eq:Bhat}).
Again we find agreement.

\section{CONCLUSIONS}
\label{sec5}

In this work we have shown that fermionic improvement works also in the Hamiltonian lattice formulation for the case of $QCD_{1+1}$.
We have computed the quark condensate and
the flavor singlet vector mass using
the $O(a^2)$ improved Hamiltonian for quarks \cite{Luo94} 
proposed by Luo, Chen, Xu, and Jiang.
In comparison with the results corresponding to the Wilson fermions without improvement, we indeed observe significant reduction of the finite lattice spacing error $O(ra)$. By comparison with analytical results for the quark condensate and the vector mass we find good agreement. 
In our opinion, this is the first lattice study of $QCD_{1+1}$ which
gives results in the strong coupling phase (in contrast to the t'Hooft phase).
In particular we present results for different gauge groups ($N_{C}=2,3,4,5,6$).
We believe that the application of the Symanzik improvement program 
to QCD in 3+1 dimensions will be very promising.

\acknowledgments
We are grateful to Q. Chen, S. Guo, J. Li and J. Liu for useful discussions.
X.Q.L. is supported by the
National Natural Science Fund for Distinguished Young Scholars,
supplemented by the
National Natural Science Foundation of China, 
fund for international cooperation and exchange,
the Ministry of Education, 
and the Hong Kong Foundation of
the Zhongshan University Advanced Research Center. 
H.K. is grateful for support by NSERC Canada.

\vfill

\begin{figure}
% GNUPLOT: LaTeX picture
\setlength{\unitlength}{0.240900pt}
\ifx\plotpoint\undefined\newsavebox{\plotpoint}\fi
\begin{picture}(1500,900)(0,0)
\font\gnuplot=cmr10 at 10pt
\gnuplot
\sbox{\plotpoint}{\rule[-0.200pt]{0.400pt}{0.400pt}}%
\put(220.0,113.0){\rule[-0.200pt]{292.934pt}{0.400pt}}
\put(220.0,113.0){\rule[-0.200pt]{4.818pt}{0.400pt}}
\put(198,113){\makebox(0,0)[r]{0}}
\put(1416.0,113.0){\rule[-0.200pt]{4.818pt}{0.400pt}}
\put(220.0,266.0){\rule[-0.200pt]{4.818pt}{0.400pt}}
\put(198,266){\makebox(0,0)[r]{0.1}}
\put(1416.0,266.0){\rule[-0.200pt]{4.818pt}{0.400pt}}
\put(220.0,419.0){\rule[-0.200pt]{4.818pt}{0.400pt}}
\put(198,419){\makebox(0,0)[r]{0.2}}
\put(1416.0,419.0){\rule[-0.200pt]{4.818pt}{0.400pt}}
\put(220.0,571.0){\rule[-0.200pt]{4.818pt}{0.400pt}}
\put(198,571){\makebox(0,0)[r]{0.3}}
\put(1416.0,571.0){\rule[-0.200pt]{4.818pt}{0.400pt}}
\put(220.0,724.0){\rule[-0.200pt]{4.818pt}{0.400pt}}
\put(198,724){\makebox(0,0)[r]{0.4}}
\put(1416.0,724.0){\rule[-0.200pt]{4.818pt}{0.400pt}}
\put(220.0,877.0){\rule[-0.200pt]{4.818pt}{0.400pt}}
\put(198,877){\makebox(0,0)[r]{0.5}}
\put(1416.0,877.0){\rule[-0.200pt]{4.818pt}{0.400pt}}
\put(220.0,113.0){\rule[-0.200pt]{0.400pt}{4.818pt}}
\put(220,68){\makebox(0,0){1}}
\put(220.0,857.0){\rule[-0.200pt]{0.400pt}{4.818pt}}
\put(463.0,113.0){\rule[-0.200pt]{0.400pt}{4.818pt}}
\put(463,68){\makebox(0,0){1.2}}
\put(463.0,857.0){\rule[-0.200pt]{0.400pt}{4.818pt}}
\put(706.0,113.0){\rule[-0.200pt]{0.400pt}{4.818pt}}
\put(706,68){\makebox(0,0){1.4}}
\put(706.0,857.0){\rule[-0.200pt]{0.400pt}{4.818pt}}
\put(950.0,113.0){\rule[-0.200pt]{0.400pt}{4.818pt}}
\put(950,68){\makebox(0,0){1.6}}
\put(950.0,857.0){\rule[-0.200pt]{0.400pt}{4.818pt}}
\put(1193.0,113.0){\rule[-0.200pt]{0.400pt}{4.818pt}}
\put(1193,68){\makebox(0,0){1.8}}
\put(1193.0,857.0){\rule[-0.200pt]{0.400pt}{4.818pt}}
\put(1436.0,113.0){\rule[-0.200pt]{0.400pt}{4.818pt}}
\put(1436,68){\makebox(0,0){2}}
\put(1436.0,857.0){\rule[-0.200pt]{0.400pt}{4.818pt}}
\put(220.0,113.0){\rule[-0.200pt]{292.934pt}{0.400pt}}
\put(1436.0,113.0){\rule[-0.200pt]{0.400pt}{184.048pt}}
\put(220.0,877.0){\rule[-0.200pt]{292.934pt}{0.400pt}}
\put(45,495){\makebox(0,0){$\chi$}}
\put(828,23){\makebox(0,0){$1/g_{lat}^2$}}
\put(220.0,113.0){\rule[-0.200pt]{0.400pt}{184.048pt}}
\put(220,397){\raisebox{-.8pt}{\makebox(0,0){$\Diamond$}}}
\put(342,404){\raisebox{-.8pt}{\makebox(0,0){$\Diamond$}}}
\put(463,409){\raisebox{-.8pt}{\makebox(0,0){$\Diamond$}}}
\put(585,413){\raisebox{-.8pt}{\makebox(0,0){$\Diamond$}}}
\put(706,418){\raisebox{-.8pt}{\makebox(0,0){$\Diamond$}}}
\put(828,422){\raisebox{-.8pt}{\makebox(0,0){$\Diamond$}}}
\put(950,426){\raisebox{-.8pt}{\makebox(0,0){$\Diamond$}}}
\put(1071,431){\raisebox{-.8pt}{\makebox(0,0){$\Diamond$}}}
\put(1193,435){\raisebox{-.8pt}{\makebox(0,0){$\Diamond$}}}
\put(1314,438){\raisebox{-.8pt}{\makebox(0,0){$\Diamond$}}}
\put(1436,442){\raisebox{-.8pt}{\makebox(0,0){$\Diamond$}}}
\put(220,495){\makebox(0,0){$+$}}
\put(342,493){\makebox(0,0){$+$}}
\put(463,492){\makebox(0,0){$+$}}
\put(585,491){\makebox(0,0){$+$}}
\put(706,492){\makebox(0,0){$+$}}
\put(828,492){\makebox(0,0){$+$}}
\put(950,492){\makebox(0,0){$+$}}
\put(1071,493){\makebox(0,0){$+$}}
\put(1193,494){\makebox(0,0){$+$}}
\put(1314,495){\makebox(0,0){$+$}}
\put(1436,497){\makebox(0,0){$+$}}
\end{picture}

\vskip 1cm

\caption{$\chi=-\langle \bar{\psi} \psi \rangle_{sub}/(g_{lat}N_c)$ versus $1/g_{lat}^{2}$ 
for $N_C=2$ with Wilson fermions. 
Crosses: $r=0.1$, Diamonds: $r=1$.}
\label{fig1}
\end{figure}
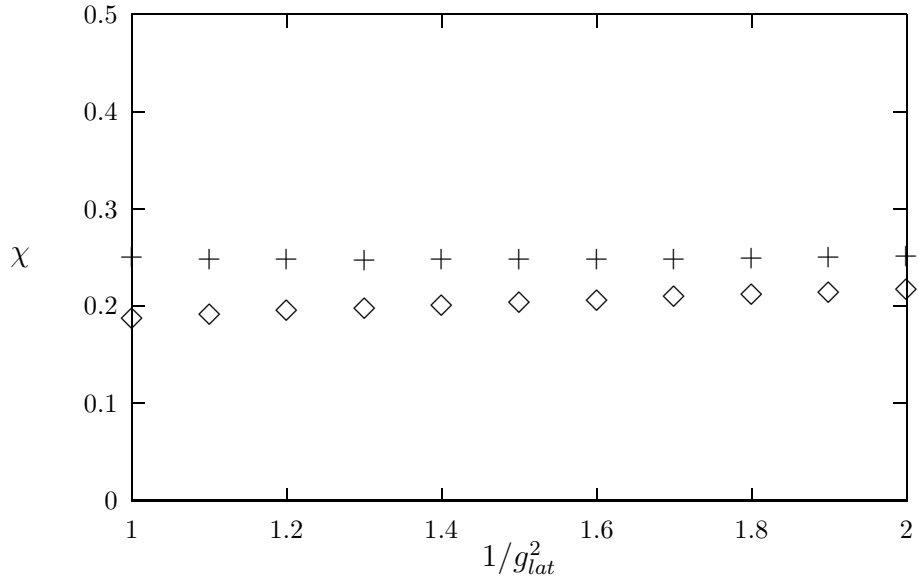

\vfill

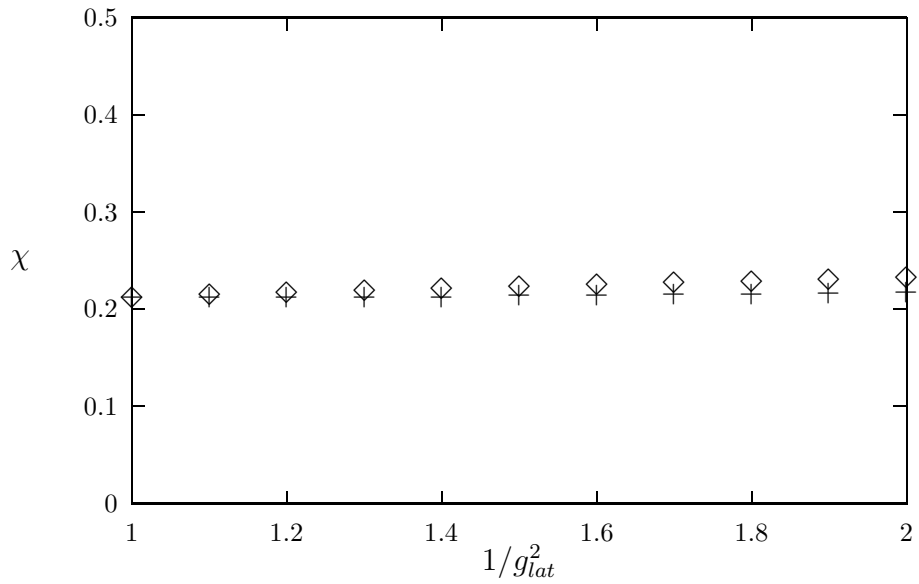
\begin{figure}
% GNUPLOT: LaTeX picture
\setlength{\unitlength}{0.240900pt}
\ifx\plotpoint\undefined\newsavebox{\plotpoint}\fi
\begin{picture}(1500,900)(0,0)
\font\gnuplot=cmr10 at 10pt
\gnuplot
\sbox{\plotpoint}{\rule[-0.200pt]{0.400pt}{0.400pt}}%
\put(220.0,113.0){\rule[-0.200pt]{292.934pt}{0.400pt}}
\put(220.0,113.0){\rule[-0.200pt]{4.818pt}{0.400pt}}
\put(198,113){\makebox(0,0)[r]{0}}
\put(1416.0,113.0){\rule[-0.200pt]{4.818pt}{0.400pt}}
\put(220.0,266.0){\rule[-0.200pt]{4.818pt}{0.400pt}}
\put(198,266){\makebox(0,0)[r]{0.1}}
\put(1416.0,266.0){\rule[-0.200pt]{4.818pt}{0.400pt}}
\put(220.0,419.0){\rule[-0.200pt]{4.818pt}{0.400pt}}
\put(198,419){\makebox(0,0)[r]{0.2}}
\put(1416.0,419.0){\rule[-0.200pt]{4.818pt}{0.400pt}}
\put(220.0,571.0){\rule[-0.200pt]{4.818pt}{0.400pt}}
\put(198,571){\makebox(0,0)[r]{0.3}}
\put(1416.0,571.0){\rule[-0.200pt]{4.818pt}{0.400pt}}
\put(220.0,724.0){\rule[-0.200pt]{4.818pt}{0.400pt}}
\put(198,724){\makebox(0,0)[r]{0.4}}
\put(1416.0,724.0){\rule[-0.200pt]{4.818pt}{0.400pt}}
\put(220.0,877.0){\rule[-0.200pt]{4.818pt}{0.400pt}}
\put(198,877){\makebox(0,0)[r]{0.5}}
\put(1416.0,877.0){\rule[-0.200pt]{4.818pt}{0.400pt}}
\put(220.0,113.0){\rule[-0.200pt]{0.400pt}{4.818pt}}
\put(220,68){\makebox(0,0){1}}
\put(220.0,857.0){\rule[-0.200pt]{0.400pt}{4.818pt}}
\put(463.0,113.0){\rule[-0.200pt]{0.400pt}{4.818pt}}
\put(463,68){\makebox(0,0){1.2}}
\put(463.0,857.0){\rule[-0.200pt]{0.400pt}{4.818pt}}
\put(706.0,113.0){\rule[-0.200pt]{0.400pt}{4.818pt}}
\put(706,68){\makebox(0,0){1.4}}
\put(706.0,857.0){\rule[-0.200pt]{0.400pt}{4.818pt}}
\put(950.0,113.0){\rule[-0.200pt]{0.400pt}{4.818pt}}
\put(950,68){\makebox(0,0){1.6}}
\put(950.0,857.0){\rule[-0.200pt]{0.400pt}{4.818pt}}
\put(1193.0,113.0){\rule[-0.200pt]{0.400pt}{4.818pt}}
\put(1193,68){\makebox(0,0){1.8}}
\put(1193.0,857.0){\rule[-0.200pt]{0.400pt}{4.818pt}}
\put(1436.0,113.0){\rule[-0.200pt]{0.400pt}{4.818pt}}
\put(1436,68){\makebox(0,0){2}}
\put(1436.0,857.0){\rule[-0.200pt]{0.400pt}{4.818pt}}
\put(220.0,113.0){\rule[-0.200pt]{292.934pt}{0.400pt}}
\put(1436.0,113.0){\rule[-0.200pt]{0.400pt}{184.048pt}}
\put(220.0,877.0){\rule[-0.200pt]{292.934pt}{0.400pt}}
\put(45,495){\makebox(0,0){$\chi$}}
\put(828,23){\makebox(0,0){$1/g_{lat}^2$}}
\put(220.0,113.0){\rule[-0.200pt]{0.400pt}{184.048pt}}
\put(220,435){\raisebox{-.8pt}{\makebox(0,0){$\Diamond$}}}
\put(342,440){\raisebox{-.8pt}{\makebox(0,0){$\Diamond$}}}
\put(463,443){\raisebox{-.8pt}{\makebox(0,0){$\Diamond$}}}
\put(585,446){\raisebox{-.8pt}{\makebox(0,0){$\Diamond$}}}
\put(706,449){\raisebox{-.8pt}{\makebox(0,0){$\Diamond$}}}
\put(828,452){\raisebox{-.8pt}{\makebox(0,0){$\Diamond$}}}
\put(950,455){\raisebox{-.8pt}{\makebox(0,0){$\Diamond$}}}
\put(1071,458){\raisebox{-.8pt}{\makebox(0,0){$\Diamond$}}}
\put(1193,460){\raisebox{-.8pt}{\makebox(0,0){$\Diamond$}}}
\put(1314,463){\raisebox{-.8pt}{\makebox(0,0){$\Diamond$}}}
\put(1436,466){\raisebox{-.8pt}{\makebox(0,0){$\Diamond$}}}
\put(220,438){\makebox(0,0){$+$}}
\put(342,437){\makebox(0,0){$+$}}
\put(463,437){\makebox(0,0){$+$}}
\put(585,437){\makebox(0,0){$+$}}
\put(706,438){\makebox(0,0){$+$}}
\put(828,440){\makebox(0,0){$+$}}
\put(950,440){\makebox(0,0){$+$}}
\put(1071,442){\makebox(0,0){$+$}}
\put(1193,443){\makebox(0,0){$+$}}
\put(1314,444){\makebox(0,0){$+$}}
\put(1436,446){\makebox(0,0){$+$}}
\end{picture}

\vskip 1cm

\caption{$\chi=-\langle \bar{\psi} \psi \rangle_{sub}/(g_{lat}N_c)$ versus $1/g_{lat}^{2}$ 
for $N_C=2$ with improved Wilson fermions. 
Crosses: $r=0.1$, Diamonds: $r=1$.}
\label{fig2}
\end{figure}

\vfill

\begin{figure}
% GNUPLOT: LaTeX picture
\setlength{\unitlength}{0.240900pt}
\ifx\plotpoint\undefined\newsavebox{\plotpoint}\fi
\begin{picture}(1500,900)(0,0)
\font\gnuplot=cmr10 at 10pt
\gnuplot
\sbox{\plotpoint}{\rule[-0.200pt]{0.400pt}{0.400pt}}%
\put(220.0,113.0){\rule[-0.200pt]{292.934pt}{0.400pt}}
\put(220.0,113.0){\rule[-0.200pt]{4.818pt}{0.400pt}}
\put(198,113){\makebox(0,0)[r]{0}}
\put(1416.0,113.0){\rule[-0.200pt]{4.818pt}{0.400pt}}
\put(220.0,266.0){\rule[-0.200pt]{4.818pt}{0.400pt}}
\put(198,266){\makebox(0,0)[r]{0.1}}
\put(1416.0,266.0){\rule[-0.200pt]{4.818pt}{0.400pt}}
\put(220.0,419.0){\rule[-0.200pt]{4.818pt}{0.400pt}}
\put(198,419){\makebox(0,0)[r]{0.2}}
\put(1416.0,419.0){\rule[-0.200pt]{4.818pt}{0.400pt}}
\put(220.0,571.0){\rule[-0.200pt]{4.818pt}{0.400pt}}
\put(198,571){\makebox(0,0)[r]{0.3}}
\put(1416.0,571.0){\rule[-0.200pt]{4.818pt}{0.400pt}}
\put(220.0,724.0){\rule[-0.200pt]{4.818pt}{0.400pt}}
\put(198,724){\makebox(0,0)[r]{0.4}}
\put(1416.0,724.0){\rule[-0.200pt]{4.818pt}{0.400pt}}
\put(220.0,877.0){\rule[-0.200pt]{4.818pt}{0.400pt}}
\put(198,877){\makebox(0,0)[r]{0.5}}
\put(1416.0,877.0){\rule[-0.200pt]{4.818pt}{0.400pt}}
\put(220.0,113.0){\rule[-0.200pt]{0.400pt}{4.818pt}}
\put(220,68){\makebox(0,0){1}}
\put(220.0,857.0){\rule[-0.200pt]{0.400pt}{4.818pt}}
\put(463.0,113.0){\rule[-0.200pt]{0.400pt}{4.818pt}}
\put(463,68){\makebox(0,0){1.2}}
\put(463.0,857.0){\rule[-0.200pt]{0.400pt}{4.818pt}}
\put(706.0,113.0){\rule[-0.200pt]{0.400pt}{4.818pt}}
\put(706,68){\makebox(0,0){1.4}}
\put(706.0,857.0){\rule[-0.200pt]{0.400pt}{4.818pt}}
\put(950.0,113.0){\rule[-0.200pt]{0.400pt}{4.818pt}}
\put(950,68){\makebox(0,0){1.6}}
\put(950.0,857.0){\rule[-0.200pt]{0.400pt}{4.818pt}}
\put(1193.0,113.0){\rule[-0.200pt]{0.400pt}{4.818pt}}
\put(1193,68){\makebox(0,0){1.8}}
\put(1193.0,857.0){\rule[-0.200pt]{0.400pt}{4.818pt}}
\put(1436.0,113.0){\rule[-0.200pt]{0.400pt}{4.818pt}}
\put(1436,68){\makebox(0,0){2}}
\put(1436.0,857.0){\rule[-0.200pt]{0.400pt}{4.818pt}}
\put(220.0,113.0){\rule[-0.200pt]{292.934pt}{0.400pt}}
\put(1436.0,113.0){\rule[-0.200pt]{0.400pt}{184.048pt}}
\put(220.0,877.0){\rule[-0.200pt]{292.934pt}{0.400pt}}
\put(45,495){\makebox(0,0){$\chi$}}
\put(828,23){\makebox(0,0){$1/g_{lat}^2$}}
\put(220.0,113.0){\rule[-0.200pt]{0.400pt}{184.048pt}}
\put(220,455){\raisebox{-.8pt}{\makebox(0,0){$\Diamond$}}}
\put(342,462){\raisebox{-.8pt}{\makebox(0,0){$\Diamond$}}}
\put(463,468){\raisebox{-.8pt}{\makebox(0,0){$\Diamond$}}}
\put(585,472){\raisebox{-.8pt}{\makebox(0,0){$\Diamond$}}}
\put(706,478){\raisebox{-.8pt}{\makebox(0,0){$\Diamond$}}}
\put(828,482){\raisebox{-.8pt}{\makebox(0,0){$\Diamond$}}}
\put(950,486){\raisebox{-.8pt}{\makebox(0,0){$\Diamond$}}}
\put(1071,491){\raisebox{-.8pt}{\makebox(0,0){$\Diamond$}}}
\put(1193,495){\raisebox{-.8pt}{\makebox(0,0){$\Diamond$}}}
\put(1314,498){\raisebox{-.8pt}{\makebox(0,0){$\Diamond$}}}
\put(1436,502){\raisebox{-.8pt}{\makebox(0,0){$\Diamond$}}}
\put(220,653){\makebox(0,0){$+$}}
\put(342,647){\makebox(0,0){$+$}}
\put(463,640){\makebox(0,0){$+$}}
\put(585,635){\makebox(0,0){$+$}}
\put(706,632){\makebox(0,0){$+$}}
\put(828,628){\makebox(0,0){$+$}}
\put(950,626){\makebox(0,0){$+$}}
\put(1071,623){\makebox(0,0){$+$}}
\put(1193,621){\makebox(0,0){$+$}}
\put(1314,621){\makebox(0,0){$+$}}
\put(1436,620){\makebox(0,0){$+$}}
\end{picture}

\vskip 1cm

\caption{$\chi=-\langle \bar{\psi} \psi \rangle_{sub}/(g_{lat}N_c)$ 
versus $1/g_{lat}^{2}$ 
for $N_C=3$ with Wilson fermions. 
Crosses: $r=0.1$, Diamonds: $r=1$.}
\label{fig3}
\end{figure}
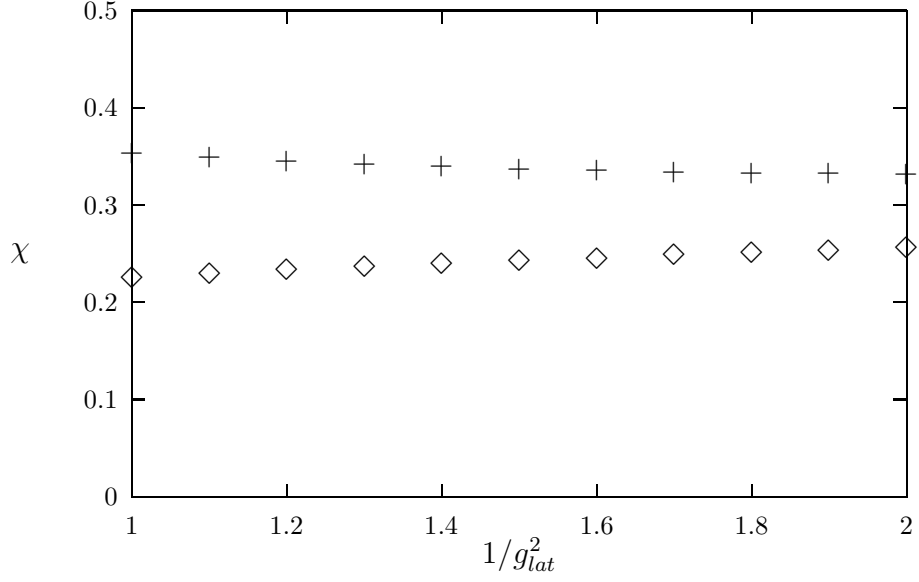

\vfill

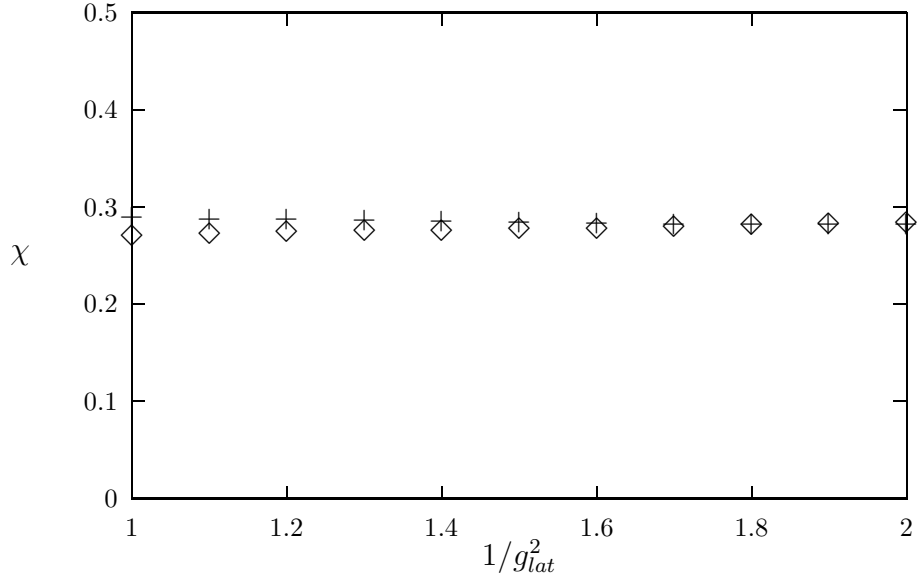
\begin{figure}
% GNUPLOT: LaTeX picture
\setlength{\unitlength}{0.240900pt}
\ifx\plotpoint\undefined\newsavebox{\plotpoint}\fi
\begin{picture}(1500,900)(0,0)
\font\gnuplot=cmr10 at 10pt
\gnuplot
\sbox{\plotpoint}{\rule[-0.200pt]{0.400pt}{0.400pt}}%
\put(220.0,113.0){\rule[-0.200pt]{292.934pt}{0.400pt}}
\put(220.0,113.0){\rule[-0.200pt]{4.818pt}{0.400pt}}
\put(198,113){\makebox(0,0)[r]{0}}
\put(1416.0,113.0){\rule[-0.200pt]{4.818pt}{0.400pt}}
\put(220.0,266.0){\rule[-0.200pt]{4.818pt}{0.400pt}}
\put(198,266){\makebox(0,0)[r]{0.1}}
\put(1416.0,266.0){\rule[-0.200pt]{4.818pt}{0.400pt}}
\put(220.0,419.0){\rule[-0.200pt]{4.818pt}{0.400pt}}
\put(198,419){\makebox(0,0)[r]{0.2}}
\put(1416.0,419.0){\rule[-0.200pt]{4.818pt}{0.400pt}}
\put(220.0,571.0){\rule[-0.200pt]{4.818pt}{0.400pt}}
\put(198,571){\makebox(0,0)[r]{0.3}}
\put(1416.0,571.0){\rule[-0.200pt]{4.818pt}{0.400pt}}
\put(220.0,724.0){\rule[-0.200pt]{4.818pt}{0.400pt}}
\put(198,724){\makebox(0,0)[r]{0.4}}
\put(1416.0,724.0){\rule[-0.200pt]{4.818pt}{0.400pt}}
\put(220.0,877.0){\rule[-0.200pt]{4.818pt}{0.400pt}}
\put(198,877){\makebox(0,0)[r]{0.5}}
\put(1416.0,877.0){\rule[-0.200pt]{4.818pt}{0.400pt}}
\put(220.0,113.0){\rule[-0.200pt]{0.400pt}{4.818pt}}
\put(220,68){\makebox(0,0){1}}
\put(220.0,857.0){\rule[-0.200pt]{0.400pt}{4.818pt}}
\put(463.0,113.0){\rule[-0.200pt]{0.400pt}{4.818pt}}
\put(463,68){\makebox(0,0){1.2}}
\put(463.0,857.0){\rule[-0.200pt]{0.400pt}{4.818pt}}
\put(706.0,113.0){\rule[-0.200pt]{0.400pt}{4.818pt}}
\put(706,68){\makebox(0,0){1.4}}
\put(706.0,857.0){\rule[-0.200pt]{0.400pt}{4.818pt}}
\put(950.0,113.0){\rule[-0.200pt]{0.400pt}{4.818pt}}
\put(950,68){\makebox(0,0){1.6}}
\put(950.0,857.0){\rule[-0.200pt]{0.400pt}{4.818pt}}
\put(1193.0,113.0){\rule[-0.200pt]{0.400pt}{4.818pt}}
\put(1193,68){\makebox(0,0){1.8}}
\put(1193.0,857.0){\rule[-0.200pt]{0.400pt}{4.818pt}}
\put(1436.0,113.0){\rule[-0.200pt]{0.400pt}{4.818pt}}
\put(1436,68){\makebox(0,0){2}}
\put(1436.0,857.0){\rule[-0.200pt]{0.400pt}{4.818pt}}
\put(220.0,113.0){\rule[-0.200pt]{292.934pt}{0.400pt}}
\put(1436.0,113.0){\rule[-0.200pt]{0.400pt}{184.048pt}}
\put(220.0,877.0){\rule[-0.200pt]{292.934pt}{0.400pt}}
\put(45,495){\makebox(0,0){$\chi$}}
\put(828,23){\makebox(0,0){$1/g_{lat}^2$}}
\put(220.0,113.0){\rule[-0.200pt]{0.400pt}{184.048pt}}
\put(220,524){\raisebox{-.8pt}{\makebox(0,0){$\Diamond$}}}
\put(342,528){\raisebox{-.8pt}{\makebox(0,0){$\Diamond$}}}
\put(463,531){\raisebox{-.8pt}{\makebox(0,0){$\Diamond$}}}
\put(585,532){\raisebox{-.8pt}{\makebox(0,0){$\Diamond$}}}
\put(706,533){\raisebox{-.8pt}{\makebox(0,0){$\Diamond$}}}
\put(828,536){\raisebox{-.8pt}{\makebox(0,0){$\Diamond$}}}
\put(950,536){\raisebox{-.8pt}{\makebox(0,0){$\Diamond$}}}
\put(1071,538){\raisebox{-.8pt}{\makebox(0,0){$\Diamond$}}}
\put(1193,541){\raisebox{-.8pt}{\makebox(0,0){$\Diamond$}}}
\put(1314,543){\raisebox{-.8pt}{\makebox(0,0){$\Diamond$}}}
\put(1436,545){\raisebox{-.8pt}{\makebox(0,0){$\Diamond$}}}
\put(220,555){\makebox(0,0){$+$}}
\put(342,553){\makebox(0,0){$+$}}
\put(463,552){\makebox(0,0){$+$}}
\put(585,551){\makebox(0,0){$+$}}
\put(706,549){\makebox(0,0){$+$}}
\put(828,548){\makebox(0,0){$+$}}
\put(950,546){\makebox(0,0){$+$}}
\put(1071,545){\makebox(0,0){$+$}}
\put(1193,545){\makebox(0,0){$+$}}
\put(1314,545){\makebox(0,0){$+$}}
\put(1436,545){\makebox(0,0){$+$}}
\end{picture}

\vskip 1cm

\caption{$\chi=-\langle \bar{\psi} \psi \rangle_{sub}/(g_{lat}N_c)$ versus $1/g_{lat}^{2}$ 
for $N_C=3$ with improved Wilson fermions. 
Crosses: $r=0.1$, Diamonds: $r=1$.}
\label{fig4}
\end{figure}

\vfill

\begin{figure}
% GNUPLOT: LaTeX picture
\setlength{\unitlength}{0.240900pt}
\ifx\plotpoint\undefined\newsavebox{\plotpoint}\fi
\begin{picture}(1500,900)(0,0)
\font\gnuplot=cmr10 at 10pt
\gnuplot
\sbox{\plotpoint}{\rule[-0.200pt]{0.400pt}{0.400pt}}%
\put(220.0,113.0){\rule[-0.200pt]{292.934pt}{0.400pt}}
\put(220.0,113.0){\rule[-0.200pt]{4.818pt}{0.400pt}}
\put(198,113){\makebox(0,0)[r]{0}}
\put(1416.0,113.0){\rule[-0.200pt]{4.818pt}{0.400pt}}
\put(220.0,304.0){\rule[-0.200pt]{4.818pt}{0.400pt}}
\put(198,304){\makebox(0,0)[r]{0.4}}
\put(1416.0,304.0){\rule[-0.200pt]{4.818pt}{0.400pt}}
\put(220.0,495.0){\rule[-0.200pt]{4.818pt}{0.400pt}}
\put(198,495){\makebox(0,0)[r]{0.8}}
\put(1416.0,495.0){\rule[-0.200pt]{4.818pt}{0.400pt}}
\put(220.0,686.0){\rule[-0.200pt]{4.818pt}{0.400pt}}
\put(198,686){\makebox(0,0)[r]{1.2}}
\put(1416.0,686.0){\rule[-0.200pt]{4.818pt}{0.400pt}}
\put(220.0,877.0){\rule[-0.200pt]{4.818pt}{0.400pt}}
\put(198,877){\makebox(0,0)[r]{1.6}}
\put(1416.0,877.0){\rule[-0.200pt]{4.818pt}{0.400pt}}
\put(220.0,113.0){\rule[-0.200pt]{0.400pt}{4.818pt}}
\put(220,68){\makebox(0,0){1}}
\put(220.0,857.0){\rule[-0.200pt]{0.400pt}{4.818pt}}
\put(463.0,113.0){\rule[-0.200pt]{0.400pt}{4.818pt}}
\put(463,68){\makebox(0,0){1.2}}
\put(463.0,857.0){\rule[-0.200pt]{0.400pt}{4.818pt}}
\put(706.0,113.0){\rule[-0.200pt]{0.400pt}{4.818pt}}
\put(706,68){\makebox(0,0){1.4}}
\put(706.0,857.0){\rule[-0.200pt]{0.400pt}{4.818pt}}
\put(950.0,113.0){\rule[-0.200pt]{0.400pt}{4.818pt}}
\put(950,68){\makebox(0,0){1.6}}
\put(950.0,857.0){\rule[-0.200pt]{0.400pt}{4.818pt}}
\put(1193.0,113.0){\rule[-0.200pt]{0.400pt}{4.818pt}}
\put(1193,68){\makebox(0,0){1.8}}
\put(1193.0,857.0){\rule[-0.200pt]{0.400pt}{4.818pt}}
\put(1436.0,113.0){\rule[-0.200pt]{0.400pt}{4.818pt}}
\put(1436,68){\makebox(0,0){2}}
\put(1436.0,857.0){\rule[-0.200pt]{0.400pt}{4.818pt}}
\put(220.0,113.0){\rule[-0.200pt]{292.934pt}{0.400pt}}
\put(1436.0,113.0){\rule[-0.200pt]{0.400pt}{184.048pt}}
\put(220.0,877.0){\rule[-0.200pt]{292.934pt}{0.400pt}}
\put(45,495){\makebox(0,0){$aM_V/g_{lat}$}}
\put(828,23){\makebox(0,0){$1/g_{lat}^2$}}
\put(220.0,113.0){\rule[-0.200pt]{0.400pt}{184.048pt}}
\put(220,483){\raisebox{-.8pt}{\makebox(0,0){$\Diamond$}}}
\put(463,495){\raisebox{-.8pt}{\makebox(0,0){$\Diamond$}}}
\put(706,507){\raisebox{-.8pt}{\makebox(0,0){$\Diamond$}}}
\put(950,519){\raisebox{-.8pt}{\makebox(0,0){$\Diamond$}}}
\put(1193,531){\raisebox{-.8pt}{\makebox(0,0){$\Diamond$}}}
\put(1436,542){\raisebox{-.8pt}{\makebox(0,0){$\Diamond$}}}
\put(220,371){\makebox(0,0){$+$}}
\put(463,383){\makebox(0,0){$+$}}
\put(706,392){\makebox(0,0){$+$}}
\put(950,399){\makebox(0,0){$+$}}
\put(1193,406){\makebox(0,0){$+$}}
\put(1436,412){\makebox(0,0){$+$}}
\end{picture}

\vskip 1cm

\caption{$aM_{V}/g_{lat}$ versus $1/g_{lat}^{2}$ 
for $N_C=2$ with Wilson fermions. 
Crosses: $r=0.1$, Diamonds: $r=1$.}
\label{fig5}
\end{figure}
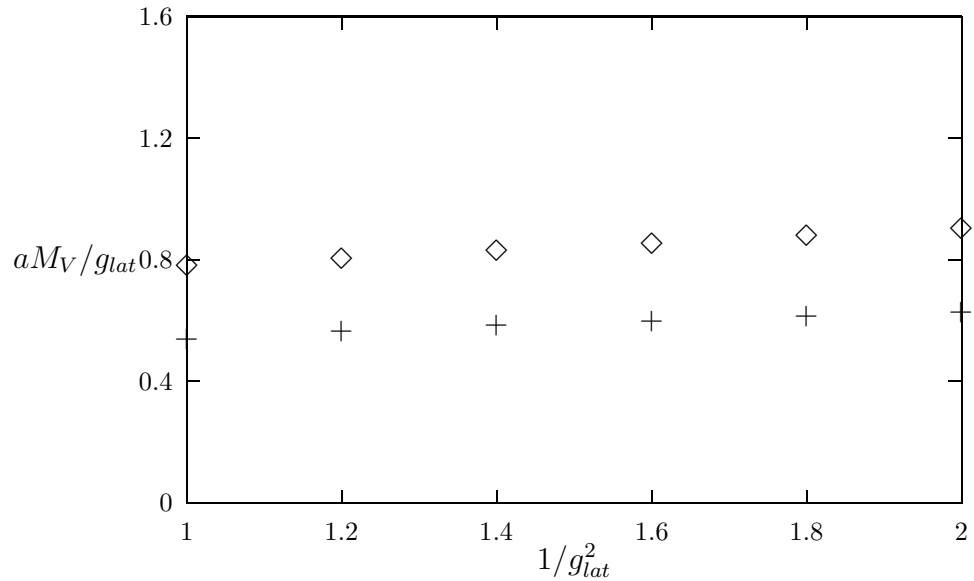

\vfill

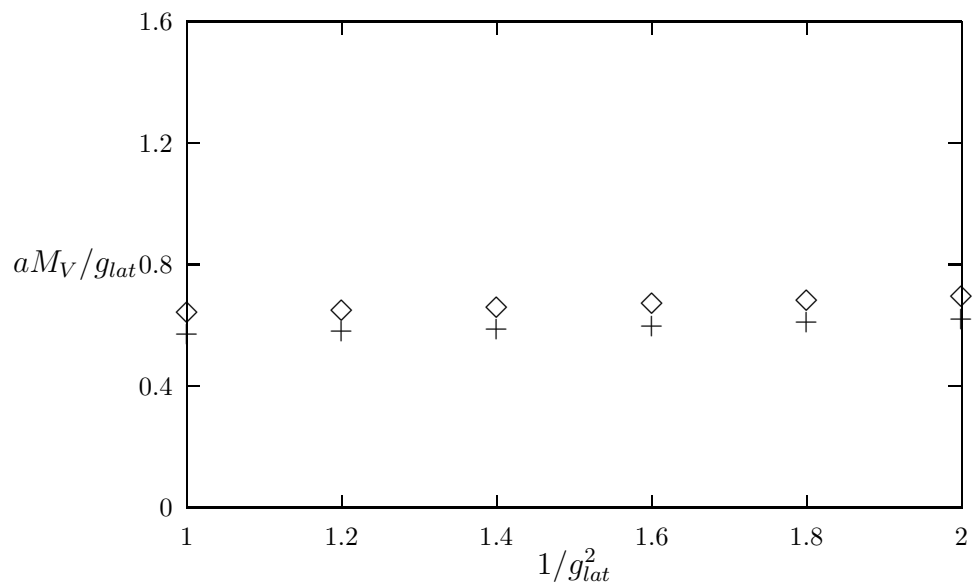
\begin{figure}

% GNUPLOT: LaTeX picture
\setlength{\unitlength}{0.240900pt}
\ifx\plotpoint\undefined\newsavebox{\plotpoint}\fi
\begin{picture}(1500,900)(0,0)
\font\gnuplot=cmr10 at 10pt
\gnuplot
\sbox{\plotpoint}{\rule[-0.200pt]{0.400pt}{0.400pt}}%
\put(220.0,113.0){\rule[-0.200pt]{292.934pt}{0.400pt}}
\put(220.0,113.0){\rule[-0.200pt]{4.818pt}{0.400pt}}
\put(198,113){\makebox(0,0)[r]{0}}
\put(1416.0,113.0){\rule[-0.200pt]{4.818pt}{0.400pt}}
\put(220.0,304.0){\rule[-0.200pt]{4.818pt}{0.400pt}}
\put(198,304){\makebox(0,0)[r]{0.4}}
\put(1416.0,304.0){\rule[-0.200pt]{4.818pt}{0.400pt}}
\put(220.0,495.0){\rule[-0.200pt]{4.818pt}{0.400pt}}
\put(198,495){\makebox(0,0)[r]{0.8}}
\put(1416.0,495.0){\rule[-0.200pt]{4.818pt}{0.400pt}}
\put(220.0,686.0){\rule[-0.200pt]{4.818pt}{0.400pt}}
\put(198,686){\makebox(0,0)[r]{1.2}}
\put(1416.0,686.0){\rule[-0.200pt]{4.818pt}{0.400pt}}
\put(220.0,877.0){\rule[-0.200pt]{4.818pt}{0.400pt}}
\put(198,877){\makebox(0,0)[r]{1.6}}
\put(1416.0,877.0){\rule[-0.200pt]{4.818pt}{0.400pt}}
\put(220.0,113.0){\rule[-0.200pt]{0.400pt}{4.818pt}}
\put(220,68){\makebox(0,0){1}}
\put(220.0,857.0){\rule[-0.200pt]{0.400pt}{4.818pt}}
\put(463.0,113.0){\rule[-0.200pt]{0.400pt}{4.818pt}}
\put(463,68){\makebox(0,0){1.2}}
\put(463.0,857.0){\rule[-0.200pt]{0.400pt}{4.818pt}}
\put(706.0,113.0){\rule[-0.200pt]{0.400pt}{4.818pt}}
\put(706,68){\makebox(0,0){1.4}}
\put(706.0,857.0){\rule[-0.200pt]{0.400pt}{4.818pt}}
\put(950.0,113.0){\rule[-0.200pt]{0.400pt}{4.818pt}}
\put(950,68){\makebox(0,0){1.6}}
\put(950.0,857.0){\rule[-0.200pt]{0.400pt}{4.818pt}}
\put(1193.0,113.0){\rule[-0.200pt]{0.400pt}{4.818pt}}
\put(1193,68){\makebox(0,0){1.8}}
\put(1193.0,857.0){\rule[-0.200pt]{0.400pt}{4.818pt}}
\put(1436.0,113.0){\rule[-0.200pt]{0.400pt}{4.818pt}}
\put(1436,68){\makebox(0,0){2}}
\put(1436.0,857.0){\rule[-0.200pt]{0.400pt}{4.818pt}}
\put(220.0,113.0){\rule[-0.200pt]{292.934pt}{0.400pt}}
\put(1436.0,113.0){\rule[-0.200pt]{0.400pt}{184.048pt}}
\put(220.0,877.0){\rule[-0.200pt]{292.934pt}{0.400pt}}
\put(45,495){\makebox(0,0){$aM_V/g_{lat}$}}
\put(828,23){\makebox(0,0){$1/g_{lat}^2$}}
\put(220.0,113.0){\rule[-0.200pt]{0.400pt}{184.048pt}}
\put(220,417){\raisebox{-.8pt}{\makebox(0,0){$\Diamond$}}}
\put(463,421){\raisebox{-.8pt}{\makebox(0,0){$\Diamond$}}}
\put(706,426){\raisebox{-.8pt}{\makebox(0,0){$\Diamond$}}}
\put(950,431){\raisebox{-.8pt}{\makebox(0,0){$\Diamond$}}}
\put(1193,437){\raisebox{-.8pt}{\makebox(0,0){$\Diamond$}}}
\put(1436,442){\raisebox{-.8pt}{\makebox(0,0){$\Diamond$}}}
\put(220,385){\makebox(0,0){$+$}}
\put(463,390){\makebox(0,0){$+$}}
\put(706,394){\makebox(0,0){$+$}}
\put(950,399){\makebox(0,0){$+$}}
\put(1193,404){\makebox(0,0){$+$}}
\put(1436,409){\makebox(0,0){$+$}}
\end{picture}

\vskip 1cm

\caption{$aM_{V}/g_{lat}$ versus $1/g_{lat}^{2}$ 
for $N_C=2$ with improved Wilson fermions. 
Crosses: $r=0.1$, Diamonds: $r=1$.}
\label{fig6}
\end{figure}

\vfill

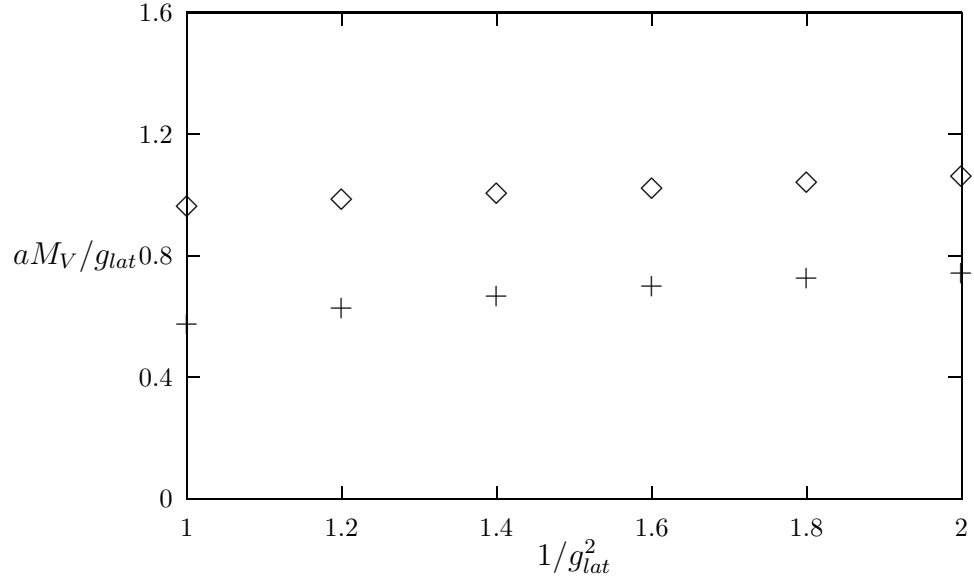
\begin{figure}
% GNUPLOT: LaTeX picture
\setlength{\unitlength}{0.240900pt}
\ifx\plotpoint\undefined\newsavebox{\plotpoint}\fi
\sbox{\plotpoint}{\rule[-0.200pt]{0.400pt}{0.400pt}}%
\begin{picture}(1500,900)(0,0)
\font\gnuplot=cmr10 at 10pt
\gnuplot
\sbox{\plotpoint}{\rule[-0.200pt]{0.400pt}{0.400pt}}%
\put(220.0,113.0){\rule[-0.200pt]{292.934pt}{0.400pt}}
\put(220.0,113.0){\rule[-0.200pt]{4.818pt}{0.400pt}}
\put(198,113){\makebox(0,0)[r]{0}}
\put(1416.0,113.0){\rule[-0.200pt]{4.818pt}{0.400pt}}
\put(220.0,304.0){\rule[-0.200pt]{4.818pt}{0.400pt}}
\put(198,304){\makebox(0,0)[r]{0.4}}
\put(1416.0,304.0){\rule[-0.200pt]{4.818pt}{0.400pt}}
\put(220.0,495.0){\rule[-0.200pt]{4.818pt}{0.400pt}}
\put(198,495){\makebox(0,0)[r]{0.8}}
\put(1416.0,495.0){\rule[-0.200pt]{4.818pt}{0.400pt}}
\put(220.0,686.0){\rule[-0.200pt]{4.818pt}{0.400pt}}
\put(198,686){\makebox(0,0)[r]{1.2}}
\put(1416.0,686.0){\rule[-0.200pt]{4.818pt}{0.400pt}}
\put(220.0,877.0){\rule[-0.200pt]{4.818pt}{0.400pt}}
\put(198,877){\makebox(0,0)[r]{1.6}}
\put(1416.0,877.0){\rule[-0.200pt]{4.818pt}{0.400pt}}
\put(220.0,113.0){\rule[-0.200pt]{0.400pt}{4.818pt}}
\put(220,68){\makebox(0,0){1}}
\put(220.0,857.0){\rule[-0.200pt]{0.400pt}{4.818pt}}
\put(463.0,113.0){\rule[-0.200pt]{0.400pt}{4.818pt}}
\put(463,68){\makebox(0,0){1.2}}
\put(463.0,857.0){\rule[-0.200pt]{0.400pt}{4.818pt}}
\put(706.0,113.0){\rule[-0.200pt]{0.400pt}{4.818pt}}
\put(706,68){\makebox(0,0){1.4}}
\put(706.0,857.0){\rule[-0.200pt]{0.400pt}{4.818pt}}
\put(950.0,113.0){\rule[-0.200pt]{0.400pt}{4.818pt}}
\put(950,68){\makebox(0,0){1.6}}
\put(950.0,857.0){\rule[-0.200pt]{0.400pt}{4.818pt}}
\put(1193.0,113.0){\rule[-0.200pt]{0.400pt}{4.818pt}}
\put(1193,68){\makebox(0,0){1.8}}
\put(1193.0,857.0){\rule[-0.200pt]{0.400pt}{4.818pt}}
\put(1436.0,113.0){\rule[-0.200pt]{0.400pt}{4.818pt}}
\put(1436,68){\makebox(0,0){2}}
\put(1436.0,857.0){\rule[-0.200pt]{0.400pt}{4.818pt}}
\put(220.0,113.0){\rule[-0.200pt]{292.934pt}{0.400pt}}
\put(1436.0,113.0){\rule[-0.200pt]{0.400pt}{184.048pt}}
\put(220.0,877.0){\rule[-0.200pt]{292.934pt}{0.400pt}}
\put(45,495){\makebox(0,0){$aM_V/g_{lat}$}}
\put(828,23){\makebox(0,0){$1/g_{lat}^2$}}
\put(220.0,113.0){\rule[-0.200pt]{0.400pt}{184.048pt}}
\put(220,571){\raisebox{-.8pt}{\makebox(0,0){$\Diamond$}}}
\put(463,581){\raisebox{-.8pt}{\makebox(0,0){$\Diamond$}}}
\put(706,590){\raisebox{-.8pt}{\makebox(0,0){$\Diamond$}}}
\put(950,599){\raisebox{-.8pt}{\makebox(0,0){$\Diamond$}}}
\put(1193,608){\raisebox{-.8pt}{\makebox(0,0){$\Diamond$}}}
\put(1436,617){\raisebox{-.8pt}{\makebox(0,0){$\Diamond$}}}
\put(220,387){\makebox(0,0){$+$}}
\put(463,412){\makebox(0,0){$+$}}
\put(706,432){\makebox(0,0){$+$}}
\put(950,447){\makebox(0,0){$+$}}
\put(1193,459){\makebox(0,0){$+$}}
\put(1436,468){\makebox(0,0){$+$}}
\end{picture}

\vskip 1cm

\caption{$aM_{V}/g_{lat}$ versus $1/g_{lat}^{2}$ 
for $N_C=3$ with Wilson fermions. 
Crosses: $r=0.1$, Diamonds: $r=1$.}
\label{fig7}

\end{figure}

\vfill

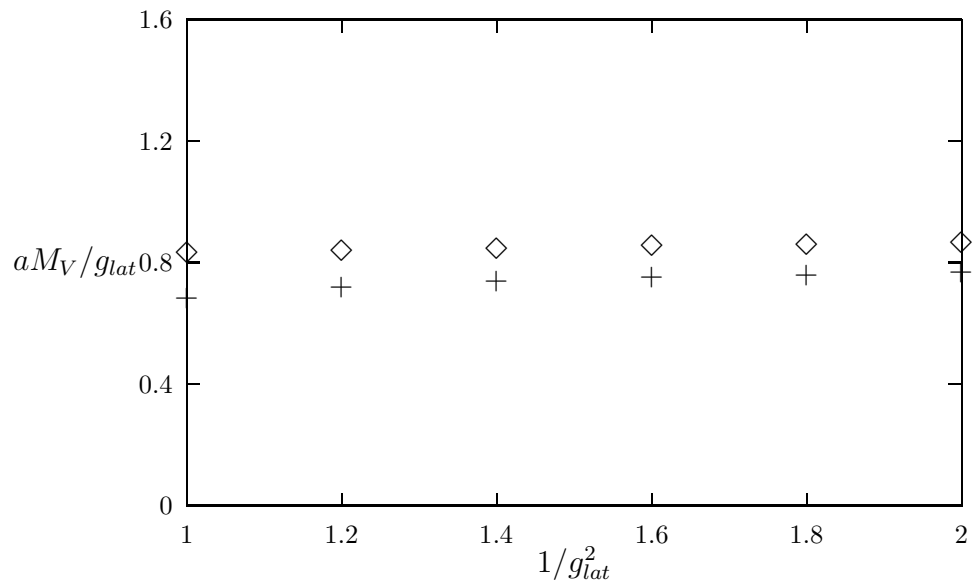
\begin{figure}
% GNUPLOT: LaTeX picture
\setlength{\unitlength}{0.240900pt}
\ifx\plotpoint\undefined\newsavebox{\plotpoint}\fi
\begin{picture}(1500,900)(0,0)
\font\gnuplot=cmr10 at 10pt
\gnuplot
\sbox{\plotpoint}{\rule[-0.200pt]{0.400pt}{0.400pt}}%
\put(220.0,113.0){\rule[-0.200pt]{292.934pt}{0.400pt}}
\put(220.0,113.0){\rule[-0.200pt]{4.818pt}{0.400pt}}
\put(198,113){\makebox(0,0)[r]{0}}
\put(1416.0,113.0){\rule[-0.200pt]{4.818pt}{0.400pt}}
\put(220.0,304.0){\rule[-0.200pt]{4.818pt}{0.400pt}}
\put(198,304){\makebox(0,0)[r]{0.4}}
\put(1416.0,304.0){\rule[-0.200pt]{4.818pt}{0.400pt}}
\put(220.0,495.0){\rule[-0.200pt]{4.818pt}{0.400pt}}
\put(198,495){\makebox(0,0)[r]{0.8}}
\put(1416.0,495.0){\rule[-0.200pt]{4.818pt}{0.400pt}}
\put(220.0,686.0){\rule[-0.200pt]{4.818pt}{0.400pt}}
\put(198,686){\makebox(0,0)[r]{1.2}}
\put(1416.0,686.0){\rule[-0.200pt]{4.818pt}{0.400pt}}
\put(220.0,877.0){\rule[-0.200pt]{4.818pt}{0.400pt}}
\put(198,877){\makebox(0,0)[r]{1.6}}
\put(1416.0,877.0){\rule[-0.200pt]{4.818pt}{0.400pt}}
\put(220.0,113.0){\rule[-0.200pt]{0.400pt}{4.818pt}}
\put(220,68){\makebox(0,0){1}}
\put(220.0,857.0){\rule[-0.200pt]{0.400pt}{4.818pt}}
\put(463.0,113.0){\rule[-0.200pt]{0.400pt}{4.818pt}}
\put(463,68){\makebox(0,0){1.2}}
\put(463.0,857.0){\rule[-0.200pt]{0.400pt}{4.818pt}}
\put(706.0,113.0){\rule[-0.200pt]{0.400pt}{4.818pt}}
\put(706,68){\makebox(0,0){1.4}}
\put(706.0,857.0){\rule[-0.200pt]{0.400pt}{4.818pt}}
\put(950.0,113.0){\rule[-0.200pt]{0.400pt}{4.818pt}}
\put(950,68){\makebox(0,0){1.6}}
\put(950.0,857.0){\rule[-0.200pt]{0.400pt}{4.818pt}}
\put(1193.0,113.0){\rule[-0.200pt]{0.400pt}{4.818pt}}
\put(1193,68){\makebox(0,0){1.8}}
\put(1193.0,857.0){\rule[-0.200pt]{0.400pt}{4.818pt}}
\put(1436.0,113.0){\rule[-0.200pt]{0.400pt}{4.818pt}}
\put(1436,68){\makebox(0,0){2}}
\put(1436.0,857.0){\rule[-0.200pt]{0.400pt}{4.818pt}}
\put(220.0,113.0){\rule[-0.200pt]{292.934pt}{0.400pt}}
\put(1436.0,113.0){\rule[-0.200pt]{0.400pt}{184.048pt}}
\put(220.0,877.0){\rule[-0.200pt]{292.934pt}{0.400pt}}
\put(45,495){\makebox(0,0){$aM_V/g_{lat}$}}
\put(828,23){\makebox(0,0){$1/g_{lat}^2$}}
\put(220.0,113.0){\rule[-0.200pt]{0.400pt}{184.048pt}}
\put(220,509){\raisebox{-.8pt}{\makebox(0,0){$\Diamond$}}}
\put(463,512){\raisebox{-.8pt}{\makebox(0,0){$\Diamond$}}}
\put(706,515){\raisebox{-.8pt}{\makebox(0,0){$\Diamond$}}}
\put(950,519){\raisebox{-.8pt}{\makebox(0,0){$\Diamond$}}}
\put(1193,522){\raisebox{-.8pt}{\makebox(0,0){$\Diamond$}}}
\put(1436,525){\raisebox{-.8pt}{\makebox(0,0){$\Diamond$}}}
\put(220,439){\makebox(0,0){$+$}}
\put(463,456){\makebox(0,0){$+$}}
\put(706,466){\makebox(0,0){$+$}}
\put(950,472){\makebox(0,0){$+$}}
\put(1193,476){\makebox(0,0){$+$}}
\put(1436,480){\makebox(0,0){$+$}}
\end{picture}

\vskip 1cm

\caption{$aM_{V}/g_{lat}$ versus $1/g_{lat}^{2}$ 
for $N_C=3$ with improved Wilson fermions. 
Crosses: $r=0.1$, Diamonds: $r=1$.}
\label{fig8}
\end{figure}

\vfill

\begin{figure}
% GNUPLOT: LaTeX picture
\setlength{\unitlength}{0.240900pt}
\ifx\plotpoint\undefined\newsavebox{\plotpoint}\fi
\sbox{\plotpoint}{\rule[-0.200pt]{0.400pt}{0.400pt}}%
\begin{picture}(1500,900)(0,0)
\font\gnuplot=cmr10 at 10pt
\gnuplot
\sbox{\plotpoint}{\rule[-0.200pt]{0.400pt}{0.400pt}}%
\put(220.0,113.0){\rule[-0.200pt]{292.934pt}{0.400pt}}
\put(220.0,113.0){\rule[-0.200pt]{0.400pt}{184.048pt}}
\put(220.0,113.0){\rule[-0.200pt]{4.818pt}{0.400pt}}
\put(198,113){\makebox(0,0)[r]{0}}
\put(1416.0,113.0){\rule[-0.200pt]{4.818pt}{0.400pt}}
\put(220.0,304.0){\rule[-0.200pt]{4.818pt}{0.400pt}}
\put(198,304){\makebox(0,0)[r]{0.1}}
\put(1416.0,304.0){\rule[-0.200pt]{4.818pt}{0.400pt}}
\put(220.0,495.0){\rule[-0.200pt]{4.818pt}{0.400pt}}
\put(198,495){\makebox(0,0)[r]{0.2}}
\put(1416.0,495.0){\rule[-0.200pt]{4.818pt}{0.400pt}}
\put(220.0,686.0){\rule[-0.200pt]{4.818pt}{0.400pt}}
\put(198,686){\makebox(0,0)[r]{0.3}}
\put(1416.0,686.0){\rule[-0.200pt]{4.818pt}{0.400pt}}
\put(220.0,877.0){\rule[-0.200pt]{4.818pt}{0.400pt}}
\put(198,877){\makebox(0,0)[r]{0.4}}
\put(1416.0,877.0){\rule[-0.200pt]{4.818pt}{0.400pt}}
\put(220.0,113.0){\rule[-0.200pt]{0.400pt}{4.818pt}}
\put(220,68){\makebox(0,0){0}}
\put(220.0,857.0){\rule[-0.200pt]{0.400pt}{4.818pt}}
\put(524.0,113.0){\rule[-0.200pt]{0.400pt}{4.818pt}}
\put(524,68){\makebox(0,0){0.05}}
\put(524.0,857.0){\rule[-0.200pt]{0.400pt}{4.818pt}}
\put(828.0,113.0){\rule[-0.200pt]{0.400pt}{4.818pt}}
\put(828,68){\makebox(0,0){0.1}}
\put(828.0,857.0){\rule[-0.200pt]{0.400pt}{4.818pt}}
\put(1132.0,113.0){\rule[-0.200pt]{0.400pt}{4.818pt}}
\put(1132,68){\makebox(0,0){0.15}}
\put(1132.0,857.0){\rule[-0.200pt]{0.400pt}{4.818pt}}
\put(1436.0,113.0){\rule[-0.200pt]{0.400pt}{4.818pt}}
\put(1436,68){\makebox(0,0){0.2}}
\put(1436.0,857.0){\rule[-0.200pt]{0.400pt}{4.818pt}}
\put(220.0,113.0){\rule[-0.200pt]{292.934pt}{0.400pt}}
\put(1436.0,113.0){\rule[-0.200pt]{0.400pt}{184.048pt}}
\put(220.0,877.0){\rule[-0.200pt]{292.934pt}{0.400pt}}
\put(45,495){\makebox(0,0){$\rho$}}
\put(828,23){\makebox(0,0){$1/N_C^2$}}
\put(220.0,113.0){\rule[-0.200pt]{0.400pt}{184.048pt}}
\put(896,424){\raisebox{-.8pt}{\makebox(0,0){$\Diamond$}}}
\put(600,434){\raisebox{-.8pt}{\makebox(0,0){$\Diamond$}}}
\put(463,441){\raisebox{-.8pt}{\makebox(0,0){$\Diamond$}}}
\put(389,445){\raisebox{-.8pt}{\makebox(0,0){$\Diamond$}}}
\put(896.0,417.0){\rule[-0.200pt]{0.400pt}{3.373pt}}
\put(886.0,417.0){\rule[-0.200pt]{4.818pt}{0.400pt}}
\put(886.0,431.0){\rule[-0.200pt]{4.818pt}{0.400pt}}
\put(600.0,417.0){\rule[-0.200pt]{0.400pt}{8.191pt}}
\put(590.0,417.0){\rule[-0.200pt]{4.818pt}{0.400pt}}
\put(590.0,451.0){\rule[-0.200pt]{4.818pt}{0.400pt}}
\put(463.0,415.0){\rule[-0.200pt]{0.400pt}{12.527pt}}
\put(453.0,415.0){\rule[-0.200pt]{4.818pt}{0.400pt}}
\put(453.0,467.0){\rule[-0.200pt]{4.818pt}{0.400pt}}
\put(389.0,411.0){\rule[-0.200pt]{0.400pt}{16.381pt}}
\put(379.0,411.0){\rule[-0.200pt]{4.818pt}{0.400pt}}
\put(379.0,479.0){\rule[-0.200pt]{4.818pt}{0.400pt}}
\sbox{\plotpoint}{\rule[-0.400pt]{0.800pt}{0.800pt}}%
\put(220,424){\usebox{\plotpoint}}
\put(220.0,424.0){\rule[-0.400pt]{292.934pt}{0.800pt}}
\end{picture}

\vskip 1cm

\caption{$\rho=-\langle \bar{\psi} \psi \rangle_{cont}/(gN_C^{3/2})$ 
in the continuum
versus $1/N_C^{2}$.
The error bars are estimated from the data for
different $r$. The full line gives the analytical result from 
Ref.\protect \cite{kn:Zhit85}.}
\end{figure}
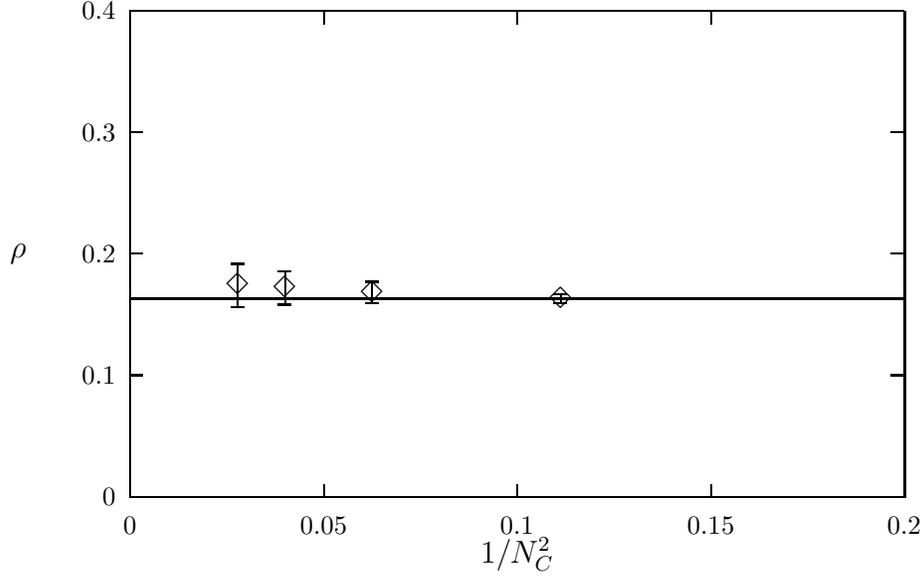
\label{fig9}
\vfill

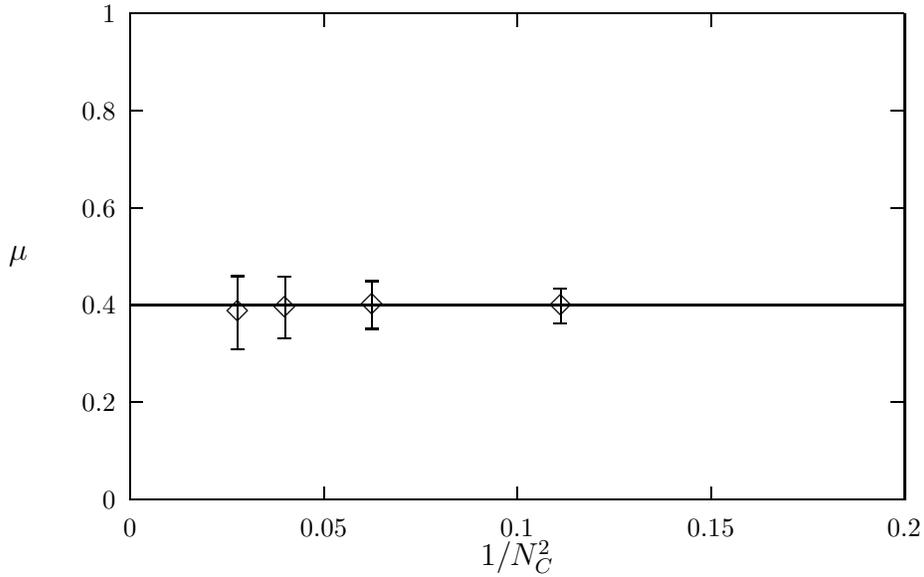
\begin{figure}
% GNUPLOT: LaTeX picture
\setlength{\unitlength}{0.240900pt}
\ifx\plotpoint\undefined\newsavebox{\plotpoint}\fi
\begin{picture}(1500,900)(0,0)
\font\gnuplot=cmr10 at 10pt
\gnuplot
\sbox{\plotpoint}{\rule[-0.200pt]{0.400pt}{0.400pt}}%
\put(220.0,113.0){\rule[-0.200pt]{292.934pt}{0.400pt}}
\put(220.0,113.0){\rule[-0.200pt]{0.400pt}{184.048pt}}
\put(220.0,113.0){\rule[-0.200pt]{4.818pt}{0.400pt}}
\put(198,113){\makebox(0,0)[r]{0}}
\put(1416.0,113.0){\rule[-0.200pt]{4.818pt}{0.400pt}}
\put(220.0,266.0){\rule[-0.200pt]{4.818pt}{0.400pt}}
\put(198,266){\makebox(0,0)[r]{0.2}}
\put(1416.0,266.0){\rule[-0.200pt]{4.818pt}{0.400pt}}
\put(220.0,419.0){\rule[-0.200pt]{4.818pt}{0.400pt}}
\put(198,419){\makebox(0,0)[r]{0.4}}
\put(1416.0,419.0){\rule[-0.200pt]{4.818pt}{0.400pt}}
\put(220.0,571.0){\rule[-0.200pt]{4.818pt}{0.400pt}}
\put(198,571){\makebox(0,0)[r]{0.6}}
\put(1416.0,571.0){\rule[-0.200pt]{4.818pt}{0.400pt}}
\put(220.0,724.0){\rule[-0.200pt]{4.818pt}{0.400pt}}
\put(198,724){\makebox(0,0)[r]{0.8}}
\put(1416.0,724.0){\rule[-0.200pt]{4.818pt}{0.400pt}}
\put(220.0,877.0){\rule[-0.200pt]{4.818pt}{0.400pt}}
\put(198,877){\makebox(0,0)[r]{1}}
\put(1416.0,877.0){\rule[-0.200pt]{4.818pt}{0.400pt}}
\put(220.0,113.0){\rule[-0.200pt]{0.400pt}{4.818pt}}
\put(220,68){\makebox(0,0){0}}
\put(220.0,857.0){\rule[-0.200pt]{0.400pt}{4.818pt}}
\put(524.0,113.0){\rule[-0.200pt]{0.400pt}{4.818pt}}
\put(524,68){\makebox(0,0){0.05}}
\put(524.0,857.0){\rule[-0.200pt]{0.400pt}{4.818pt}}
\put(828.0,113.0){\rule[-0.200pt]{0.400pt}{4.818pt}}
\put(828,68){\makebox(0,0){0.1}}
\put(828.0,857.0){\rule[-0.200pt]{0.400pt}{4.818pt}}
\put(1132.0,113.0){\rule[-0.200pt]{0.400pt}{4.818pt}}
\put(1132,68){\makebox(0,0){0.15}}
\put(1132.0,857.0){\rule[-0.200pt]{0.400pt}{4.818pt}}
\put(1436.0,113.0){\rule[-0.200pt]{0.400pt}{4.818pt}}
\put(1436,68){\makebox(0,0){0.2}}
\put(1436.0,857.0){\rule[-0.200pt]{0.400pt}{4.818pt}}
\put(220.0,113.0){\rule[-0.200pt]{292.934pt}{0.400pt}}
\put(1436.0,113.0){\rule[-0.200pt]{0.400pt}{184.048pt}}
\put(220.0,877.0){\rule[-0.200pt]{292.934pt}{0.400pt}}
\put(45,495){\makebox(0,0){$\mu$}}
\put(828,23){\makebox(0,0){$1/N_C^2$}}
\put(220.0,113.0){\rule[-0.200pt]{0.400pt}{184.048pt}}
\put(896,417){\raisebox{-.8pt}{\makebox(0,0){$\Diamond$}}}
\put(600,418){\raisebox{-.8pt}{\makebox(0,0){$\Diamond$}}}
\put(463,414){\raisebox{-.8pt}{\makebox(0,0){$\Diamond$}}}
\put(389,407){\raisebox{-.8pt}{\makebox(0,0){$\Diamond$}}}
\put(896.0,390.0){\rule[-0.200pt]{0.400pt}{13.009pt}}
\put(886.0,390.0){\rule[-0.200pt]{4.818pt}{0.400pt}}
\put(886.0,444.0){\rule[-0.200pt]{4.818pt}{0.400pt}}
\put(600.0,381.0){\rule[-0.200pt]{0.400pt}{18.067pt}}
\put(590.0,381.0){\rule[-0.200pt]{4.818pt}{0.400pt}}
\put(590.0,456.0){\rule[-0.200pt]{4.818pt}{0.400pt}}
\put(463.0,366.0){\rule[-0.200pt]{0.400pt}{23.367pt}}
\put(453.0,366.0){\rule[-0.200pt]{4.818pt}{0.400pt}}
\put(453.0,463.0){\rule[-0.200pt]{4.818pt}{0.400pt}}
\put(389.0,349.0){\rule[-0.200pt]{0.400pt}{27.703pt}}
\put(379.0,349.0){\rule[-0.200pt]{4.818pt}{0.400pt}}
\put(379.0,464.0){\rule[-0.200pt]{4.818pt}{0.400pt}}
\sbox{\plotpoint}{\rule[-0.400pt]{0.800pt}{0.800pt}}%
\put(220,418){\usebox{\plotpoint}}
\put(220.0,418.0){\rule[-0.400pt]{292.934pt}{0.800pt}}
\end{picture}

\vskip 1cm

\caption{$\mu=M_{V}/(g \sqrt{N_{C}+1})$ in the continuum versus $1/N_C^{2}$.
The error bars are estimated  from the data for
different $r$. The full line gives the analytical result from 
Ref.\protect \cite{kn:Bhat82}.}
\label{fig10}
\end{figure}

\end{document}